\documentclass[
10pt,aps,
notitlepage, 
bibnotes,
prb,
twocolumn,
showpacs,preprintnumbers,superscriptaddress,
amsmath,amssymb,floatfix]{revtex4-1}

\usepackage{graphicx}
\usepackage{dsfont}
\usepackage{dcolumn}
\usepackage{bm}
\usepackage{color}
\usepackage{xcolor}
\usepackage{url}
\usepackage{tabularx}
\usepackage{array}
\usepackage{booktabs}
\usepackage{comment}
\usepackage{hyperref}
\usepackage{footnote}
\usepackage{lipsum}
\usepackage[normalem]{ulem}
\usepackage{booktabs}
\usepackage{esint}

\DeclareSymbolFont{md}{OMX}{mdput}{m}{n} 
\DeclareMathSymbol{\intop}{\mathop}{md}{90}

\definecolor{colorA}{rgb}{0, 0, 1}
\definecolor{colorB}{rgb}{0.5, 0, 0.9}
\definecolor{colorC}{rgb}{0.4, 0, 0.4}
\definecolor{color_green}{rgb}{0, 0.39, 0}

\newcommand{\BBS}[1]{{\color{color_green} #1}} 

\definecolor{pacificb}{HTML}{1CA9C9}

\makeatletter
  \def\my@tag@font{\normalsize}
  \def\maketag@@@#1{\hbox{\m@th\normalfont\my@tag@font#1}}
  \let\amsmath@eqref\eqref
  \renewcommand\eqref[1]{{\let\my@tag@font\relax\amsmath@eqref{#1}}}
\makeatother

\begin{document}

\title{Symmetry-Governed Dynamics of Magnetic Skyrmions Under Field Pulses}

\author{Vladyslav~M.~Kuchkin}
\email{vladyslav.kuchkin@ulu}
\affiliation{Department of Physics and Materials Science, University of Luxembourg, L-1511 Luxembourg, Luxembourg}

\author{Bruno~Barton-Singer}
\affiliation{Institute of Applied and Computational Mathematics, Foundation for Research and Technology - Hellas, 700 13 Heraklion, Greece}

\author{Pavel~F.~Bessarab}
\affiliation{Science Institute, University of Iceland, 107 Reykjavík, Iceland}
\affiliation{Department of Physics and Electrical Engineering, Linnaeus University, SE-39231 Kalmar, Sweden}

\author{Nikolai~S.~Kiselev}
 \affiliation{Peter Gr\"unberg Institute, Forschungszentrum J\"ulich and JARA, 52425 J\"ulich, Germany}

\date{\today}

\begin{abstract}
Topological magnetic solitons, such as skyrmions, exhibit intriguing particle-like properties that make them attractive for fundamental research and practical applications. While many magnetic systems can host skyrmions as statically stable configurations, chiral magnets stand out for their ability to accommodate a wide diversity of skyrmions with arbitrary topological charges and varied morphologies. Despite extensive investigation, a complete understanding of chiral magnetic skyrmions has remained elusive. We present a classification of all chiral skyrmions, demonstrating three classes based on their response to external magnetic field pulses: stationary, translating, and rotating. We highlight the role of magnetic texture symmetry in this classification. Skyrmions with varied dynamics offer avenues for exploring phenomena like skyrmion-skyrmion scattering that might be crucial for future applications.
\end{abstract}
\maketitle

\vspace{0.25cm}
\noindent

\noindent
Skyrmions are well-localized in space, topologically nontrivial magnetic textures, attractive for future information computing devices, from classical~\cite{Tokura2020} to neuromorphic~\cite{Song_20} and quantum~\cite{Panagopoulos_21}.
Skyrmions were first predicted ~\cite{Bogdanov_89} and experimentally observed~\cite{Muhlbauer_09,Yu_10} in chiral magnets -- a special type of materials characterized by competing Heisenberg exchange and Dzyaloshinskii-Moria interaction~\cite{Dzyaloshinskii,Moriya} (DMI).
The prominent examples of chiral magnets are B20-type crystals, e.g., Fe$_{1-x}$Co$_x$Si~\cite{Yu_10, Park_14}, FeGe~\cite{Yu_11, Kovacs_17, Du_18, Yu_18}, MnSi~\cite{Muhlbauer_09,Yu_15}, and others compounds~\cite{Shibata_13}. 
Besides that, skyrmions were also observed in many other magnetic crystals and heterostructures~\cite{Tokura2020}.

The whole family of chiral magnets can be roughly divided into three-dimensional (3D) and two-dimensional (2D) systems.
Excluding apparent phenomena unique to 3D systems, e.g., the skyrmion braiding effect~\cite{Zheng_21} or stability of chiral bobbers~\cite{Zheng_18} and hopfion rings~\cite{Zheng_23}, many phenomena can occur both in 2D and 3D systems.
A prominent example of such phenomena is the coexistence of skyrmions and antiskyrmions, which first was theoretically predicted by 2D model~\cite{Kuchkin_20i} and later were observed by means of Lorentz TEM in a thin film of FeGe~\cite{Zheng_22}.
As such, the predictive capacity of the 2D model in chiral magnets provides a robust framework for anticipating effects across both 2D and 3D systems.

In the micromagnetic framework, the total energy of a thin plate of an isotropic chiral magnet can be written as follows:
\begin{equation}
\mathcal{E}=\int \{\mathcal{A}\left|\nabla\mathbf{n}\right|^{2} + \mathcal{D} w_\mathrm{D}(\textbf{n})-M_\mathrm{s}\mathbf{B}_\mathrm{e} \cdot\mathbf{n}\}\,  \Delta l\,\mathrm{d}S,
\label{Etot}
\end{equation}
where $\mathbf{n}$ is the magnetization unit vector field, 
$M_\mathrm{s}$ is the saturation magnetization of the material, 
$\mathcal{A}$ and $\mathcal{D}$ are the exchange stiffness constant and the DMI constant, respectively. 
Since we consider the 2D model, the magnetization field $\mathbf{n}$ is assumed uniform across the film thickness, $\Delta l$.

The DMI term $w_\mathrm{D}(\mathbf{n})$ 
is defined by combinations of Lifshitz invariants, $\Lambda_{ij}^{(k)}\!=\! n_i\partial_k n_j\!-\!n_j\partial_k n_i$.
Without loss of generality, in our calculations, we assume Bloch-type DMI where $w(\mathbf{n})\! =  \!\Lambda_{zy}^{(x)}\!+\! \Lambda_{xz}^{(y)}\!$.
However, the results presented here are also valid for systems with N\'{e}el-type modulations~\cite{Romming_13, Kez_15, Romming_15} as well as for crystals with D$_{2\mathrm{d}}$ or S$_4$ point group symmetry~\cite{Kosuke_21, Tang_23}.

The last term in \eqref{Etot} is the Zeeman interaction with the external magnetic field, which is assumed perpendicular to the plane of the film, $\mathbf{B}_\mathrm{ext}=B\mathbf{e}_\mathrm{z}$.
In a more particular case, the Hamiltonian \eqref{Etot} can also include the magnetocrystalline anisotropy and the demagnetizing field energy terms. 
These terms, however, do not change the presented results qualitatively, so 
we exclude them in favor of the simplicity.
Moreover, we employ dimensionless units of distances given with respect to the equilibrium period of chiral modulations at the ground state, $L_\mathrm{D}=4\pi\mathcal{A}/\mathcal{D}$.
The strength of the external magnetic field is given in units of the saturation field, $B_\mathrm{D}=\mathcal{D}^2/(2 M_\mathrm{s}\mathcal{A})$.

\begin{figure*}[ht]
\centering
\includegraphics[width=18cm]{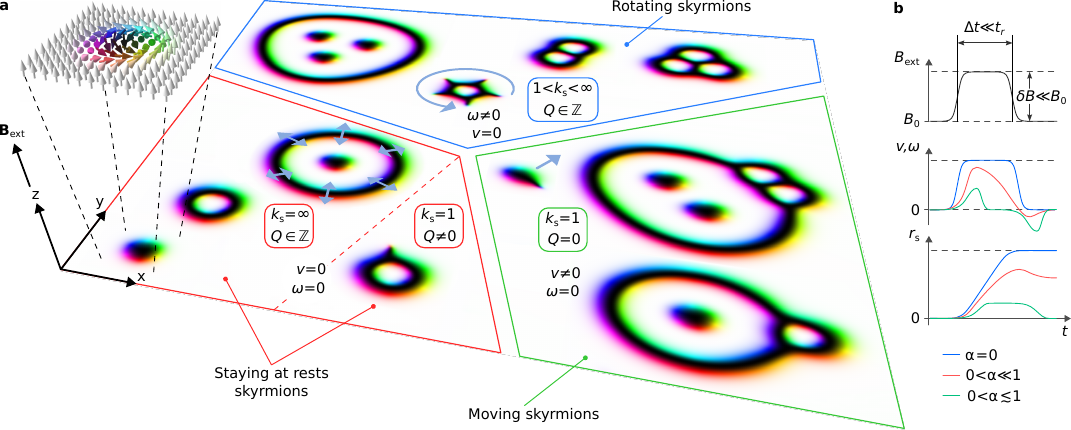}
\caption{\small
\textbf{Three fundamental classes of chiral magnetic skyrmions}.
\textbf{a} shows diverse solutions of chiral magnetic solitons in 2D that can be decomposed into three classes based on their response to a step-like pulse of the external magnetic field.
The order of the skyrmion rotation symmetry $(k_\mathrm{s})$ and skyrmion topological charge $Q$ uniquely define the skyrmion dynamical class.
Topologically trivial solutions with low symmetry, $Q=0$ and $k_\mathrm{s}=1$, belong to the class of solitons moving with constant linear velocity, $v\neq 0$.
The high symmetry skyrmions with $1<k_\mathrm{s}<\infty$ are the solitons rotating about their geometrical center with constant angular velocity, $\omega\neq 0$.
All other skyrmions belong to the class of solitons, which exhibit the breathing mode only, with $v = 0$ and $\omega = 0$.
The top left inset shows the magnetic vector field of a $\pi$-skyrmion schematically and explains the standard color code used throughout the paper.
\textbf{b} shows the magnetic field pulse $\delta B(t)$ and the dynamical response of the solitons: velocity $(v,\omega)$ and collective coordinate $r_\mathrm{s}$ as a function of time.
Dependencies $v,\omega,r _\mathrm{s}$ on time are different for various dissipation regimes: zero damping (blue), infinitesimally small damping (red), and strong damping (green).
\label{Fig1}
}
\end{figure*}

Recently, a series of works~\cite{Rybakov_19, Foster_19, Kuchkin_20ii, Barton-Singer_20, Kuchkin_23, Barton-Singer_23} reported the discovery of a wide diversity of magnetic skyrmions as statically stable solutions of the 2D model of chiral magnet. 
In Fig.~\ref{Fig1}\textbf{a}, we provide representative examples of such solutions, including skyrmion bags~\cite{Rybakov_19, Foster_19} and skyrmions with chiral kinks~\cite{Kuchkin_20ii}. 
So-called tailed skyrmions, which were reported recently~\cite{Kuchkin_23}, are not shown in Fig.~\ref{Fig1}\textbf{a} but are also discussed in this study.

Until now, the dynamical properties of chiral magnetic skyrmions with arbitrary topological charge have been discussed mainly in the context of transnational motion induced by electric current~\cite{Zeng_20, Kuchkin_21, Kind_21}.
It was shown that besides the topological charge,
\begin{equation}
    Q = \dfrac{1}{4\pi}\int\mathbf{n}\cdot\left[\partial_\mathrm{x}\mathbf{n}\times\partial_\mathrm{y}\mathbf{n}\right] \mathrm{d}S,
\end{equation}
which defines the skyrmion Hall effect, the rotational symmetry of skyrmions represents another important characteristic determining the dynamics~\cite{Kuchkin_21}.
The order of rotational symmetry $k_\mathrm{s}$ of a 2D localized magnetic texture is defined as the number of times the texture is transformed into itself when it undergoes a complete in-plane rotation. 
In other words, $k_\mathrm{s}$ is the maximum integer for which the following symmetry holds~\cite{Kuchkin_21}:
\begin{equation}
\mathbf{n} (\mathbf{r}) = \mathcal{R}(\pm \phi)\,\mathbf{n} \!\left(\mathcal{R}(-\phi)\mathbf{r}\right), \label{Rn}
\end{equation}
where $\mathcal{R}(\phi)$ is the $3\times3$ rotation matrix 
about the $z$-axis by angle $\phi=2\pi/k_\mathrm{s}$. 
The sign $\pm$ in Eq.~\eqref{Rn} accounts for different types of DMI [see Ref.~\cite{Kuchkin_21} for details]. 
In the case of axially symmetric skyrmions, e.g., ordinary $\pi$-skyrmion and skyrmionium, one has
$k_\mathrm{s}=\infty$.

Here, we show that the whole family of 2D chiral magnetic skyrmions can be categorized into three classes based on skyrmion's topological index, $Q$, and its order of rotational symmetry, $k_\mathrm{s}$ [Fig.~\ref{Fig1}\textbf{a}].
Each class is characterized by a particular dynamics in response to a short pulse of external magnetic field.  
As we show below performing numerical simulations with the Landau-Lifshitz-Gilbert (LLG) equation, some skyrmions exhibit translational motion, some
rotational motion and some skyrmions stay at rest
without motion.

\vspace{0.5cm}
\noindent
\textbf{Results}

\noindent
\textbf{External field pulse}.
\noindent
It is well known that the motion of magnetic skyrmions can be induced by various stimuli, e.g., by a gradient of the internal parameters (for instance, $\mathcal{A}$, $\mathcal{D}$ or $M_\mathrm{s}$) or by external parameters such as applied magnetic field~\cite{Malozemoff_79}, or temperature~\cite{Zang_13}.
The translation motion of magnetic skyrmions also can be induced by electrical current~\cite{SpinTransferTorques, Zang_11}, magnon waves~\cite{Schutte_14}, etc.
Here we consider the skyrmion dynamics induced by the short pulses of the perpendicular external magnetic field, $B_\mathrm{ext}(t)=B_{0}+\delta B(t)$,  where $B_{0}$ is a constant magnetic field, and $\delta B(t)\ll B_{0}<B_\mathrm{D}$ is a small perturbation varying over time [Fig.~\ref{Fig1}\textbf{b}]. 
In the case of zero damping, $\alpha=0$, the whole energy the pulse is transferred to the magnetic texture dynamics, which, depending on the skyrmion type can correspond to translational or rotational motion of the skyrmion [Fig.~\ref{Fig1}\textbf{b}].
For $\delta B(t)\ll B_\mathrm{D}$, the translational or rotational velocity follows the same time dependence as $\delta B(t)$ [blue lines in Fig.~\ref{Fig1}\textbf{b}].
In this regime, the skyrmion is set in motion after the first increasing pulse and returns to rest after the second decaying pulse.
Between the pulses, the skyrmion velocity remains constant.

In the case of finite damping, $\alpha>0$, the behavior of the system will depend on the ratio between the interval between the increasing and decreasing pulses $\Delta t$ and the relaxation time, $t_\mathrm{r}$, which in the first approximation is inversely proportional to the damping parameter, $t_\mathrm{r}\sim\alpha^{-1}$.
For a short interval between pulses, $\Delta t \ll t_\mathrm{r}$, when the system is not able to reach relaxation, the velocity of the skyrmion might exhibit the sign flip for a short period of time.
As in the case of zero damping, the resulting distance, $ r_\mathrm{s}$, traveled by the skyrmion in this case, will still be non-zero [red lines in Fig.~\ref{Fig1}\textbf{b}].
Thereby, two cases of $\alpha=0$ and $0<\alpha\ll 1$ are qualitatively identical.

On the contrary, in the case of  $\Delta t \gtrsim t_\mathrm{r}$, [green lines in Fig.~\ref{Fig1}\textbf{b}], the skyrmion reaches relaxation between the pulses and returns to its initial position after application of the second pulse.
This scenario, which may naturally take place in the system with high damping, leads to the vibration of skyrmions -- when the position of the skyrmion is changing in time, but the average distance traveled by skyrmion equals zero, $ r_\mathrm{s}=0$.
Note that all of the above is valid not only for translation but also for the rotational dynamics of skyrmions.

The results presented in this work are based on the nondissipative dynamics approach, which, as we have mentioned above, remain qualitatively equivalent to the case of small damping, $0<\alpha\ll 1$, and short pulses.
Numerical simulations of such skyrmion dynamics were performed in Mumax with Gilbert damping, $\alpha$, set to zero.
To ensure that the whole energy of the pulse is transferred into skyrmion motion, we use pulses of a smooth profile and very weak intensity, $\delta B=0.01 B_\mathrm{D}$.
Otherwise, the field pulse can excite combined vibrational translational motion modes, which may even lead to skyrmion instability.
In most cases, we succeed in decoupling two types of excitations. However, in some cases, it proved impossible to completely avoid the coupling between translational and vibrational modes.
Such cases are discussed separately in the text.

\vspace{0.5cm}
\noindent
\textbf{Symmetry analysis}.
The skyrmion behavior observed in the numerical experiment and summarised in Fig.~\ref{Fig1} can be understood using the collective coordinate approach. 
In the literature, this approach, based on the assumption of rigid translation or rotation of magnetic solitons, is also known as the Thiele approach~\cite{Thiele_73}. 
For further convenience, we parametrize magnetization using the spherical angles, $\mathbf{n}=(\sin\Theta\cos\Phi,\sin\Theta\sin\Phi,\cos\Theta)$
and consider translational and rotational motion of skyrmions separately.

\begin{figure*}
\centering
\includegraphics[width=17.7cm]{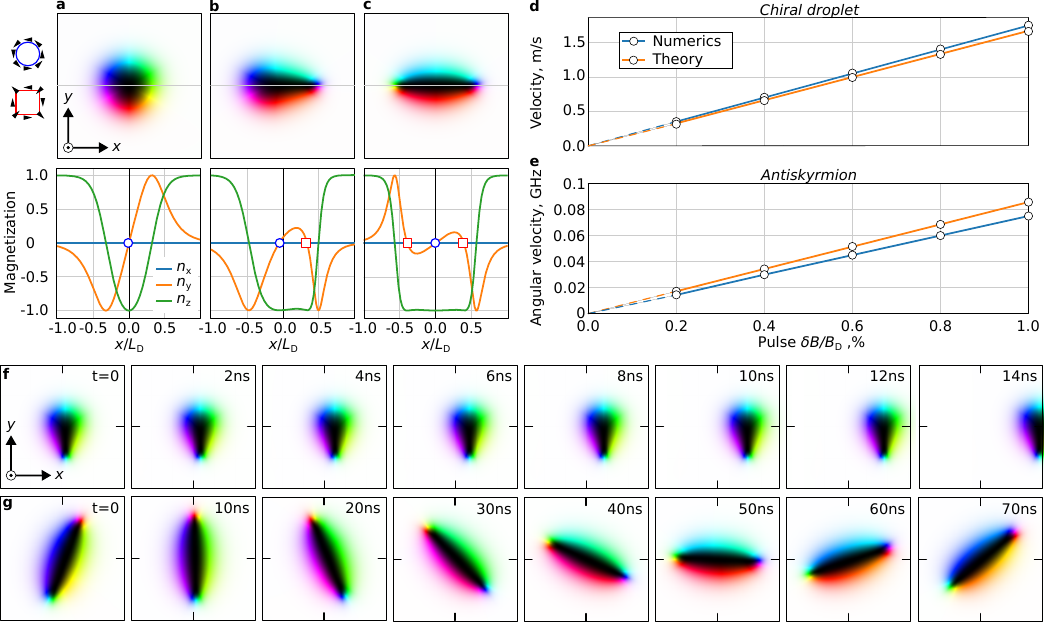}
\caption{\small \textbf{The most compact skyrmions of the three classes}.
\textbf{a}, \textbf{b} and \textbf{c} show magnetic textures of the $\pi$-skyrmion, chiral droplet, and antiskyrmion stabilized at $h=0.62$.
The magnetization components $\mathbf{n}(x)$ along the line $y=0$ are shown below. Hollow circles and squares denote the position of solitons' cores $n_\mathrm{x}=n_\mathrm{y}=0$ and $n_\mathrm{z}=-1$ with winding $\nu=+1$ and $\nu=-1$, respectively.
\textbf{d} and \textbf{e} show the comparison between velocities obtained in simulations and found analytically for the chiral droplet and antiskyrmion for the pulses of various amplitude.
Dynamics for those solitons excited by the pulse $\delta B=0.01 B_\mathrm{D}$ can be seen from \textbf{f} and \textbf{g} providing the system snapshots at different times.
}
\label{Fig4}
\end{figure*}

In the case of translational motion induced by external field pulse, the corresponding equation of motion can be written as $\Phi(\mathbf{r},t)=\Phi(\mathbf{r}-\boldsymbol{v}t)$ and  $\Theta(\mathbf{r},t)=\Theta(\mathbf{r}-\boldsymbol{v}t)$ where $\boldsymbol{v}=(v_{\mathrm{x}}, v_{\mathrm{y}})$ is a slowly-varying in time velocity of skyrmion.
The velocity $\boldsymbol{v}$ should satisfy the Thiele equation~\cite{Thiele_73}:
\begin{equation}
	\mathbf{e}_\mathrm{z}\times\boldsymbol{v}\int q\mathrm{d}S =0,
	\label{Thiele_v}
\end{equation}
where $q=(4\pi)^{-1}\left(\partial_\mathrm{x}\Theta\partial_\mathrm{y}\Phi-\partial_\mathrm{y}\Theta\partial_\mathrm{x}\Phi\right)\sin\Theta$ is the topological charge density.
It is worth emphasizing that the Thiele equation is derived from the energy conservation law.

In the case of rotational motion, the dynamics of magnetic texture can be described as $\Phi(x,y,t)=\Phi(x^\prime,y^\prime)-\omega t$ and  $\Theta(x,y,t)=\Theta(x^\prime,y^\prime)$, where $\omega$ is the angular velocity and $(x^\prime,y^\prime)$ are rotated coordinates:
\begin{eqnarray}
	&& x^{\prime}=x\cos \omega t-y\sin\omega t,\,\,
        y^{\prime}=y\cos \omega t+x\sin \omega t.
\end{eqnarray}
In this case, the Thiele equation can be written as~\cite{Papanicolaou}:
\begin{equation}
	\omega\!\int q x\mathrm{d}S =0,\,\,\omega\!\int q y\mathrm{d}S =0.
	\label{Thiele_w}
\end{equation}

Interestingly, equations \eqref{Thiele_v} and \eqref{Thiele_w} can only yield non-trivial solutions ($\boldsymbol{v}\neq 0$ or $\omega\neq 0$) when the corresponding integrals of $q$, $qx$, and $qy$ are zero.
In the next sections, we provide the exact equations for these velocities by considering an approach that goes beyond the Thiele approximation.
However, to distinguish magnetic skyrmions by their dynamics from the very general perspective, it is enough to consider Eqs.~\eqref{Thiele_v} and \eqref{Thiele_w}.
In particular, assuming a skyrmion satisfies either Eq.\eqref{Thiele_v} or Eq.\eqref{Thiele_w}, we can categorize skyrmions into three types based on their rotational symmetry order, $k_\mathrm{s}$, and topological charge, $Q$.

Let us start with Eq.~\eqref{Thiele_w} for skyrmion rotation. 
The charge density $q$ has the same rotation symmetry as the corresponding magnetization field, and can be written as a Fourier series in polar coordinates $(x,y)=(\rho\cos\phi, \rho\sin\phi)$.
It is easy to show that $q\sim a\cos(k_\mathrm{s}\phi)+b\sin(k_\mathrm{s}\phi)$ and thus the necessary condition of zeroing integrals $qx$ and $qy$ in ~\eqref{Thiele_w} is $k_\mathrm{s}>1$.
Thereby, low-symmetry skyrmions with $k_\mathrm{s}=1$ cannot rotate with fixed angular velocity because, in this case, Eq.~\eqref{Thiele_w} has only one solution, $\omega=0$. 
On the other hand, any high symmetry skyrmions with $k_\mathrm{s}\geq 2$ possess the ability to rotate.
The aforementioned criterion applies to skyrmions of any arbitrary topological charge, including topologically trivial solitons.
Finally, it is worth noting that in the case of axially symmetric solutions, $k_\mathrm{s}=\infty$, the angular velocity $\omega$, is ill-defined.
The latter is easy to understand since, for an axially symmetric soliton in a perpendicular field, there are no orientational parameters, and all directions are equivalent.
Because of that, all axially symmetric skyrmions must be excluded from the first class of solutions characterized by rotational motion.

Now let us consider the second class of solutions for skyrmions, which satisfies Eq.~\eqref{Thiele_v} for translational motion.
It is evident that to make the integral in ~\eqref{Thiele_v} equal to zero, the skyrmion must be topologically trivial, $Q=0$.
On the other hand, the solutions with $Q=0$ formally satisfy the above criteria for rotating skyrmions when they have a high symmetry.
The only solutions, which do not overlap with the first class are the topologically trivial $Q=0$, skyrmions with single-fold symmetry, $k_\mathrm{s}=1$. 
In the next section, we provide more rigorous arguments in favor of this statement.

The third class is composed of all the other skyrmions that do not satisfy the criteria for the first and second classes.
All axially symmetric solutions belong to this class irrespective of the topological charge.
Besides that, all low-symmetry skyrmions with  $k_\mathrm{s}=1$ and nonzero topological charges, $Q\neq0$, also belong to the third class.
In agreement with the above, the numerical simulations show that the magnetic field pulses cannot induce either rotation or translation of these skyrmions.
Magnetic field pulses excite only self-vibrating modes of such skyrmions, which can be effectively suppressed by using smooth profile pulses [Supplementary Movie 1].

\begin{figure*}[ht!]
\centering
\includegraphics[width=16.5cm]{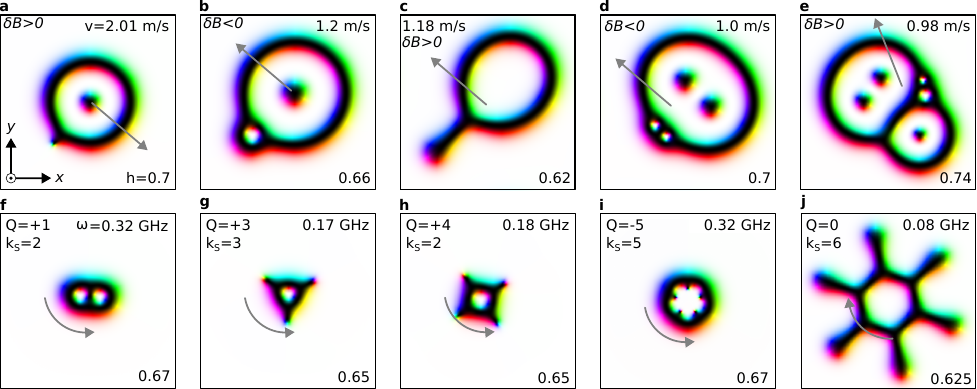}
\caption{~\small Dynamics of moving and rotating skyrmions. \textbf{a}-\textbf{e} show examples of topologically trivial skyrmions ($Q=0$) with $k_\mathrm{s}=1$. 
\textbf{f}-\textbf{j} show the magnetic skyrmions with $k_\mathrm{s}>1$. 
All states are stabilized at given magnetic fields, $h$.
Blue arrows show the direction of the motion.
The provided values of skyrmion velocities correspond to the pulse  $|\delta B|=0.01B_\mathrm{D}$.
}
\label{Fig5}
\end{figure*}

\vspace{0.5cm}
\noindent
\textbf{Skyrmion velocities}.
\noindent
The classification of chiral skyrmions outlined in the previous section is qualitative.
The latter means that the exact values for the linear, $v$, and angular, $\omega$, skyrmion velocities,  do not follow from the Thiele equations~\eqref{Thiele_v}, \eqref{Thiele_w}. 
Deriving analytical expressions for velocities and their dependencies on the model parameters, e.g., pulse intensity, $\delta B$, requires going beyond the Thiele approach.
One way of doing that is to take into account the conservation of other quantities rather than energy conservation as in the original Thiele approach.
Earlier, Papanicolaou and Tomaras showed~\cite{Papanicolaou} that at zero damping, a linear momentum conservation law holds for moving solitons and a angular momentum conservation law holds for rotating solitons.
In Supplementary Note 1, we derive the following equations for translational skyrmion motion:
\begin{equation}
    \boldsymbol{v}\cdot\boldsymbol{p}=-\gamma m\delta{B},
    \label{velocity_eq_v}
\end{equation}
and for skyrmion rotation:
\begin{equation}
    \omega(l+m)=-\gamma m\delta{B}.
    \label{velocity_eq_w}
\end{equation}
In \eqref{velocity_eq_v} and \eqref{velocity_eq_w}, $\gamma$ is the gyromagnetic ratio, $m$ is remanent magnetization along the $z$-axis, $\boldsymbol{p}$ is the skyrmion linear momentum, $l$ is the skyrmion angular momentum which can be written as follows~\cite{Papanicolaou}:
\begin{eqnarray}    \boldsymbol{p}&=&\displaystyle\int\mu\left(\mathbf{e}_x\dfrac{\partial\Phi}{\partial x}+\mathbf{e}_y\dfrac{\partial\Phi}{\partial y}\right)\mathrm{d}S\label{p_int},\\
l&=&\displaystyle\int\mu\left(y\dfrac{\partial\Phi}{\partial x}-x\dfrac{\partial\Phi}{\partial y}\right)\mathrm{d}S,
\label{l_int}
\end{eqnarray}
where $\mu=(4\pi)^{-1}(1-\cos\Theta)$ is magnon density.
Since equations \eqref{velocity_eq_v} and \eqref{velocity_eq_w} are linear with respect to velocities, they are straightforward to solve.
However, the calculation of the momentum integrals, $\boldsymbol{p}$ and $l$ in \eqref{p_int}, \eqref{l_int} (Supplementary Note 2), may present a challenge.
Nevertheless, in some simple cases discussed in the following section, $\boldsymbol{p}$ and $l$ can be easily calculated.
Furthermore, one can demonstrate a connection between the momenta $\boldsymbol{p}$ and $l$ and the topological charge density $q$ [Supplementary Note 2].

One of the advantages of Eqs.~\eqref{velocity_eq_v} and \eqref{velocity_eq_w} is that contrary to the original Thiele equation, they can describe the dynamics where the velocities are allowed to change in time.
The latter, however, is true for negligibly small damping only. 
Moreover, it is important to emphasize that  Eqs.~\eqref{velocity_eq_v} and \eqref{velocity_eq_w} can be used only together with the above symmetry analysis and skyrmion classification based on Thiele equations~\eqref{Thiele_v} and \eqref{Thiele_w}.
The above statement can be illustrated as follows. 
Let us consider any axially symmetric skyrmion, which, according to the symmetry analysis of Thiele equations and numerical experiment, exhibits neither rotational ($\omega=0$) nor translational motion ($v=0$).
The formal nonzero solutions of \eqref{velocity_eq_v} and \eqref{velocity_eq_w}, in this case, do not have a physical meaning.

\vspace{0.5 cm}
\noindent
\textbf{The dynamics of the most compact skyrmions}.
As follows from \eqref{velocity_eq_v}, \eqref{velocity_eq_w},  translational $(\boldsymbol{v})$ and angular $(\omega)$ velocities are linearly proportional to the field pulse, $\delta B$. 
Proportionality coefficients represent $\boldsymbol{p}$ and $l$, the values of which are not so obvious since the corresponding integrals in \eqref{p_int}, \eqref{l_int} depend on the details of the skyrmion structure.
As it was shown in Ref.~\cite{Papanicolaou}, for skyrmions with \textit{simple cores}, the momenta $\boldsymbol{p}$ and $l$ can be calculated straightforwardly.
By \textit{simple cores}, we assume that the regions with $n_\mathrm{z}=-1$ represent points.
In the opposite case, when the regions with $n_\mathrm{z}=-1$ represent extended segments, e.g., closed lines, arcs, etc., we say that the skyrmion has \textit{complex cores}. 

The most compact skyrmions characterized by the presence of only simple cores are depicted in Fig.~\ref{Fig4} \textbf{a}-\textbf{c}.
The $\pi$-skyrmion with $Q=-1$ depicted in \textbf{a} belongs to the class of resting solitons, 
the chiral droplet~\cite{droplet} with $Q=0$ in \textbf{b} belongs to moving solitons, and
the antiskyrmion~\cite{Kuchkin_20i} with $Q=1$ in \textbf{c} belongs to the class of rotating skyrmions.
As follows from the magnetization profiles, the $\pi$-skyrmion, chiral droplet and antiskyrmion have one, two, and three simple cores, respectively.

According to our classification, only the droplet and antiskyrmion can demonstrate pulse-induced motion. 
In the numerical experiment, we estimated the velocities of the chiral droplet and antiskyrmion for various pulse strengths (Method).
In Fig.~\ref{Fig4} \textbf{d} and \textbf{e}, we provide this data together with the solutions of Eqs.~\eqref{velocity_eq_v}, \eqref{velocity_eq_w} for comparison.
We observe a good agreement for small pulse amplitude.
The snapshots of the droplet and antiskyrmion at different times are depicted in \textbf{f} and \textbf{g}, respectively. For the chosen parameters (Method), the droplet linear velocity is $\sim 1.7 m/s$, and the antiskyrmion angular velocity is $\sim 70$ MHz. 

\begin{figure*}[ht!]
\centering
\includegraphics[width=17.1cm]{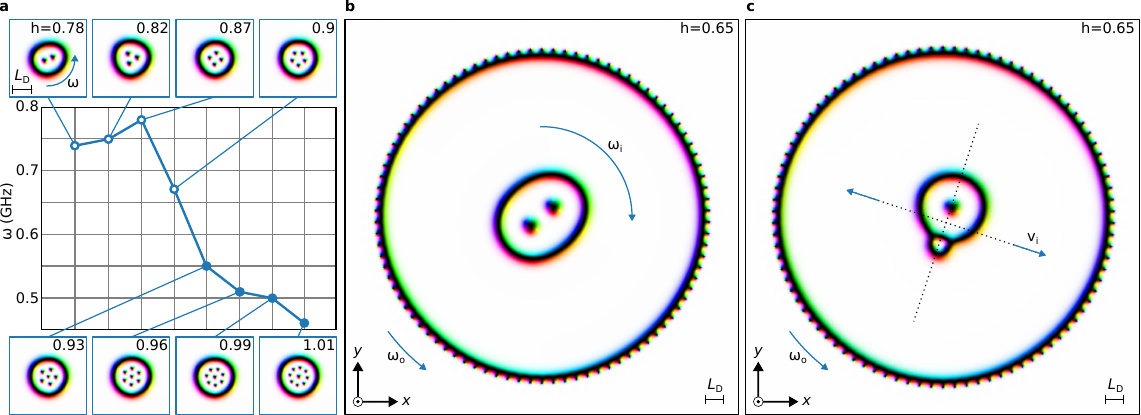}
\caption{~\small Complex rotational dynamics of skyrmion bags. \textbf{a} shows angular velocity, $\omega$, for a set of high-symmetry skyrmion bags stabilized at various magnetic fields $h=B_\mathrm{ext}/B_\mathrm{D}$. 
The field values indicated in each inset are chosen to achieve bags of nearly same sizes. 
All inset images have identical sizes of $4 L_\mathrm{D}\times 4 L_\mathrm{D}$.
\textbf{b} and \textbf{c} show nested skyrmion bags which exhibit more complex dynamics.  
In both cases, the outer bag, which contains multiple chiral kinks, rotates counterclockwise with angular velocity $\omega_\mathrm{o}$. 
The inner skyrmion bag either rotates clockwise with angular velocity $\omega_\mathrm{i}$ an in  \textbf{b} or moves with linear velocity $v_\mathrm{i}$ as in \textbf{c}.
Due to the interaction with the outer skyrmion bags, the inner skyrmion in \textbf{c} exhibits back-and-forth motion.  
}
\label{Fig7}
\end{figure*}

\noindent
\textbf{Dynamics of skyrmion bags, kinked and tailed skyrmions}. To illustrate the validity of the above classification, we examined a large variety of statically stable solutions that were predicted before~\cite{Rybakov_19, Foster_19, Kuchkin_20ii, Barton-Singer_20, Kuchkin_23}.
For instance, Figs.~\ref{Fig5} \textbf{a}-\textbf{e} show the set of topologically trivial skyrmions with single-fold symmetry, $k_\mathrm{s}=1$.
Supplementary Movie 2 illustrates the dynamics of skyrmions depicted in Figs.~\ref{Fig5}~\textbf{a}-\textbf{e} and the droplet soliton after applying the pulse.
Comparison of the velocities allows us to deduce that the $3\pi$-skyrmion with chiral kink depicted in \textbf{a} has the highest speed, $2.01$ m/s among all other skyrmions.
Note, that all textures depicted in Fig.~\ref{Fig5} are stabilized at slightly different applied fields, but their dynamics are induced by the pulse of the same amplitude, $\delta B$.

Skyrmions \textbf{a}-\textbf{d} have a \textit{mirror} symmetry with respect to line $x=y$.
The linear velocity of such skyrmions is always pointing perpendicular to the mirror axis [Supplementary Note 2].
The example of a solution with a lack of such mirror symmetry is given in \textbf{e}.

For the skyrmionium with a tail depicted in \textbf{c},
we observe the presence of additional vibrations.
We explain such vibration by the excitation of a skyrmion breathing mode. 
In the case of other skyrmions, the presence of such vibrations is much less prominent and can be thought of as a small perturbation on top of the main type of dynamics.
The excitation of these modes can be suppressed by applying a smoother pulse shape.
That means that the transition from $B_{0}$ to $B_{0}+\delta B$ is more extended in time,  similar to what is indicated in Supplementary Movie 1.
In the case of tailed skyrmions, a smooth shape profile may not be sufficient, and one has to reduce the pulse amplitude in addition.

\begin{figure*}
\centering
\includegraphics[width=18cm]{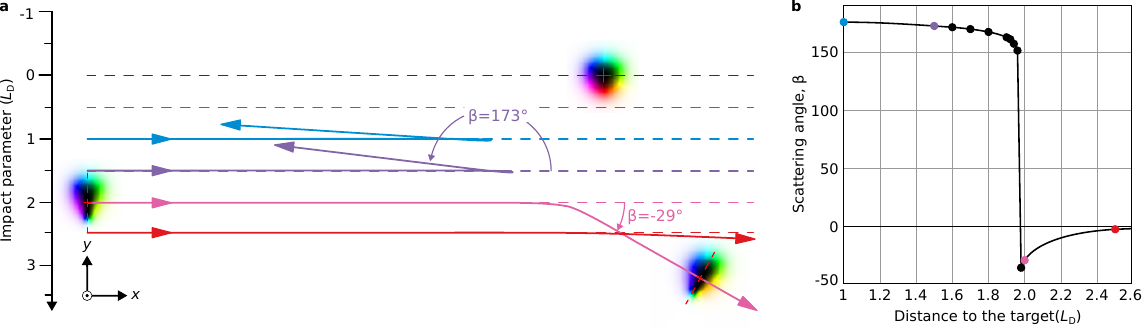}
\caption{~\small Pulse-induced skyrmion scattering. \textbf{a} shows the trajectories of the chiral droplet (bullet particle) at different distances to the $\pi$-skyrmion (target particle).
For the distance of $2L_\mathrm{D}$, we show the chiral droplet at two distinct points on its trajectory to illustrate its rotation during scattering.
Such an interaction process is characterized by the scattering angle, $\beta\in[-180^{\circ},180^{\circ}]$.
\textbf{b} shows the dependency of the angle $\beta$ for different distances between bullet and target.
The discontinuity in this dependency corresponds to a distance of $1.97 L_\mathrm{D}$.
}
\label{Fig9}
\end{figure*}

Different high-symmetry rotating skyrmions are shown in Figs.~\ref{Fig5}\textbf{f}-\textbf{j}.
The skyrmion bag in \textbf{a} and skyrmion with chiral kinks on the inner shell in \textbf{i} have the highest angular velocities, $\sim 0.32$ GHz.
Skyrmion bags with chiral kinks on the outer shell depicted in \textbf{g} and \textbf{h} rotate with almost twice smaller angular velocity, $\omega=0.17$ GHz and $0.18$ GHz, respectively.
The skyrmionium with tails depicted on \textbf{j} has the lowest angular velocity.
The dynamics of rotated skyrmions is illustrated in Supplementary Movie 3.

From results provided in Figs.~\ref{Fig5}\textbf{f}-\textbf{j} we can deduce that skyrmion bags are good candidates for the fastest-rotating skyrmions.
We studied various skyrmion bags with positive and negative topological charges to explore this conjecture more systematically.
We concluded that for achieving the maximum $\omega$, skyrmion bags with negative $Q$ are more efficient.

To keep all these skyrmion bags depicted in Fig.~\ref{Fig7} approximately the same size, one must apply different magnetic fields.
The solutions can be divided into two groups depending on the absence (open symbols) of the presence (solid symbols) of the $\pi$-skyrmion in the center of the bag. 
Although this classification is somewhat arbitrary, the angular velocities of skyrmion bags from these two groups are quite different.
For the chosen micromagnetic parameters, the skyrmion bag in \textbf{c} has the highest angular velocity $\omega=0.78$~GHz.
The rotation of skyrmion bags depicted in Fig.~\ref{Fig7} is illustrated in Supplementary Movie 4.
It is worth noting that in these simulations, we utilized high-order discretization schemes for numerical calculation of spatial derivatives in the Hamiltonian \eqref{Etot} and effective field terms in the LLG equation~\cite{Kuchkin_21}.

The wide diversity of skyrmion bags of different sizes and symmetry stabilized at identical conditions allows the study of complex dynamics beyond ordinary rotational or translational motion. 
Such complex dynamics can be observed, for instance, when one skyrmion bag is nested inside another skyrmion bag of larger size, as depicted in Fig.~\ref{Fig7}\textbf{b} and \textbf{c}.
The size of the outer skyrmion bag can be controlled by the number of attached chiral kinks, which tend to be equidistantly distributed with period $\sim1 L_\mathrm{D}$ along the perimeter.
The applied magnetic field pulse excites the rotational motion of the outer high-symmetry skyrmion bag.
When the inner skyrmion also has high-symmetry $k_\mathrm{s}>1$, as in Fig.~\ref{Fig7}\textbf{b}, it also rotates.
However, the rotation directions of the inner and outer skyrmion bags do not necessarily coincide.
Decreasing the diameter of the outer skyrmion bag, such as by reducing the number of chiral kinks and allowing closer interaction between the inner and outer skyrmion bags, will affect their angular velocities.
Below the critical size of the outer skyrmion bag, the angular velocity of the inner skyrmion can even change the sign, as illustrated in Supplementary Movie 5.

If the inner skyrmion is topologically trivial and has $k_\mathrm{s}=1$ as in Fig.~\ref{Fig7}\textbf{c}, the applied magnetic field pulse will excite its translational motion. 
Since the center of the outer skyrmion ring is fixed, at some point, the inner skyrmion starts to interact with the outer skyrmion bag. 
As a result of elastic interaction between skyrmions, the linear velocity $v_\mathrm{i}$ of the inner skyrmion changes the direction, which leads to the back-and-forth motion of the inner skyrmion bag.
Supplementary Movie 6 illustrates the dynamics of skyrmion bags depicted in Fig.~\ref{Fig7}\textbf{b} and \textbf{c}.

\noindent
\textbf{Skyrmion-skyrmion scattering}. The different dynamical response of skyrmions to the magnetic field pulse provides a unique opportunity to study skyrmion scattering process.
To illustrate this effect, in Fig.~\ref{Fig9} we show the results of micromagnetic simulations for chiral droplet and the $\pi$-skyrmion.
We select chiral droplet and $\pi$-skyrmion as most compact solitons in the class of moving and resting skyrmions, respectively.
Thereby, $\pi$-skyrmion represents a \textit{target particle}, and the chiral droplet plays the role of a \textit{bullet particle}.

In the absence of the $\pi$-skyrmion, the trajectories of the chiral droplet represent straight lines (see dashed lines in  Fig.~\ref{Fig9}\textbf{a}).
In our setup, this line is parallel to the $x$-axis.
We conducted a series of micromagnetic simulations with different \textit{impact parameter}.
While the observed interaction shares similarities with Rutherford's scattering, the functional dependence of the deflection angle on the impact parameter is distinctly different.
In particular, we found that deflection angle $\beta$ can change its sign depending on the impact parameter.
Our estimation suggests that the critical distance, where the deflection angles abruptly change sign, occurs at $\sim 2L_\mathrm{D}$.
To elucidate this deflection angle behavior, one must consider interparticle interaction potentials like those discussed in Refs.~\cite{Kuchkin_20i,Barton-Singer_23} regarding skyrmion-antiskyrmion interaction. 
However, an in-depth exploration of this topic falls outside the scope of this work. 
Our primary objective here is to introduce the external field pulse excitations as a promising approach for investigating the scattering processes of chiral skyrmions.

Even more elaborate scattering processes occur when the target particle is a soliton that can move. 
Supplementary Movie 7 illustrates the three most representative cases of skyrmion scattering.
An interesting effect occurs when a topologically trivial skyrmionium is the target particle. 
According to the above classification, the applied pulse cannot induce skyrmionium motion, but interaction with the bullet particle can. 
This statement is in agreement with Eq.~\ref{Thiele_v}.
The direction of skyrmionium motion is primarily defined by the impact parameter between the droplet and the target. 
The linear momentum conservation law holds in this case. As seen in Supplementary Movie 7, the speed of the chiral droplet decreases after partially transferring its linear momentum to the skyrmionium.

Similar to the droplet's scattering on the $\pi$-skyrmion, the scattering on a rotating skyrmion bag resembles the elastic scattering of particles. 
It is intriguing that the dynamics of magnetic skyrmions, described by first-order differential equations with respect to time, exhibit so many similarities to the dynamics of Newtonian particles, which are described by second-order differential equations with respect to time. 
On the other hand, there are many differences. 
For instance, besides the aforementioned flip of the scattering angle sign [Fig.~\ref{Fig9}], it is noticeable that the angular velocity of the skyrmion bag does not change when it interacts with the chiral droplet (see Supplementary Movie 7). 
These and other effects related to skyrmion-skyrmion scattering require a separate study, which will be presented elsewhere.

\noindent
\textbf{Conclusions}
We have studied the magnetic field pulse-induced dynamics of 2D chiral magnetic skyrmions. The presented results remain valid for arbitrary external excitations that effectively act as a magnetic field pulse, e.g., laser pulses, in zero and weak damping regimes.

It is shown that, based on their symmetry and topological index, all chiral skyrmions can be divided into three classes: breathing, moving, and rotating skyrmions. We have demonstrated that high-symmetry skyrmions can rotate with GHz-order frequency.

We have shown that skyrmion velocities can be calculated semi-analytically using the derived equations, which are consistent with the conservation laws of angular and linear momentum. Additionally, we illustrate the complex dynamics of nested skyrmion bags. 
Finally, we investigated skyrmion scattering on different targets and showed that, while there are some similarities with Rutherford scattering, skyrmion-skyrmion scattering is generally a much more complicated phenomenon.

\noindent
\textbf{Methods}

\noindent
\textbf{Micromagnetic simulations}.

Micromagnetic simulations were performed in Mumax~\cite{Mumax}. 
The standard cuboid size in our simulations was 1 nm$\times$ 1 nm$\times$1 nm.
The DMI constant is set such that the equilibrium period of helical modulations equals $L_\mathrm{D}=4\pi\mathcal{A}/\mathcal{D}=64$ nm.
All simulations are performed assuming periodical boundary conditions (PBC) in $xy$-plane.  
To increase the accuracy of the micromagnetic simulations, in some cases, we employed a fourth-order finite difference scheme in the calculation of the energy and effective field terms~\cite{ Donahue, Rybakov_19, Kuchkin_21}.
For details on implementing the finite difference scheme, see the Supplementary mumax script.

The dynamics discussed in the main text have been excited by pulses of the external magnetic field applied perpendicular to the magnetic film, $\delta B(t)$, defined as follows
\begin{equation}
    \delta B(t) = A\left(1-\dfrac{1+\exp(-\nu t_{0})}{1+\exp(\nu(t-t_{0}))}\right),\label{pulse_shape}
\end{equation}
with amplitude $A=0.01 B_\mathrm{D}$, time scaling factor $\nu=1$ GHz and time offset, $t_{0}=10$ ns. 
It is worth noting that we use a single increasing pulse.

\vspace{0.5cm}
\noindent
\textbf{Estimation of velocities from micromagnetic simulations.}
\noindent
To estimate the linear velocity of skyrmion, we trace its position, $\left(x_\mathrm{s}, y_\mathrm{s}\right)$, using the following formula optimized for the simulations with PBC~\cite{Kuchkin_21}: 
\begin{align}
    x_\mathrm{s}\!=\!\dfrac{L_\mathrm{x}}{2\pi}\mathrm{arctan}\left(\dfrac{\int\!\mathcal{N}_\mathrm{y}\sin\left(2\pi x/L_\mathrm{x}\right)\mathrm{d}x}{\int\!\mathcal{N}_\mathrm{y}\cos\left(2\pi x/L_\mathrm{x}\right)\mathrm{d}x}\right)\pm l_\mathrm{x}L_\mathrm{x},\\ 
    y_\mathrm{s}\!=\!\dfrac{L_\mathrm{y}}{2\pi}\mathrm{arctan}\left(\dfrac{\int\!\mathcal{N}_\mathrm{x}\sin\left(2\pi y/L_\mathrm{y}\right)\mathrm{d}y}{\int\!\mathcal{N}_\mathrm{x}\cos\left(2\pi y/L_\mathrm{y}\right)\mathrm{d}y}\right)\pm l_\mathrm{y}L_\mathrm{y},\label{center_mass}
\end{align}
where $\mathcal{N}_\mathrm{x}\equiv\mathcal{N}_\mathrm{x}(y)=\int(1-n_\mathrm{z})\mathrm{d}x$ and $\mathcal{N}_\mathrm{y}\equiv\mathcal{N}_\mathrm{y}(x)=\int(1-n_\mathrm{z})\mathrm{d}y$.
The integer numbers $l_\mathrm{x}$ and $l_\mathrm{y}$ account for crossing the domain boundary in the $x$ and $y$ directions, respectively.
The sign $(\pm)$ indicates the positive or negative direction of skyrmion motion along the corresponding axis.
Knowing the skyrmion's position at any moment in time, the estimation of its velocity is straightforward.

For calculation of the angular velocity, $\omega$, we first determine the number $k_\mathrm{s}$ for each soliton and then perform the Fourier transformations in polar coordinates:
\begin{eqnarray}
 & m_\mathrm{c}(t) = \intop_{0}^{2\pi}(1-\cos\Theta(r,\phi,t))\cos(k_\mathrm{s}\phi)r\mathrm{d}r\mathrm{d}\phi,\nonumber\\
 & m_\mathrm{s}(t) = \intop_{0}^{2\pi}(1-\cos\Theta(r,\phi,t))\sin(k_\mathrm{s}\phi)r\mathrm{d}r\mathrm{d}\phi.\label{Fourier_w}
\end{eqnarray}
Both functions, $m_\mathrm{c}(t)$, $m_\mathrm{s}(t)$ represent harmonic functions of period $1/\omega$ which can be employed to find $\omega$ accurately.
Due to the fact that simulations are done on the rectangular domain, integration in \eqref{Fourier_w} can be done in Cartesian coordinates. For this purpose, we rewrite trigonometric functions as follows,
\begin{eqnarray}
 & \sin(k_\mathrm{s}\phi)=\dfrac{1}{r^{k_\mathrm{s}}}\displaystyle\sum_{n=0}^{\left\lfloor(k_\mathrm{s}-1)/2\right\rfloor}(-1)^{n}C_{k_\mathrm{s}}^{2n+1}x^{k_\mathrm{s} - 2n-1}y^{2n+1},\nonumber\\
 & \cos(k_\mathrm{s}\phi) = \dfrac{1}{r^{k_\mathrm{s}}}\displaystyle\sum_{n=0}^{\left\lfloor k_\mathrm{s}/2\right\rfloor }(-1)^{n}C_{k_\mathrm{s}}^{2n}x^{k_\mathrm{s} - 2n}y^{2n},\label{sincos_xy}
\end{eqnarray}
where $C_{a}^{b}=a!/b!(a-b)!$ is the binomial coefficient.

\begin{center}
    {\footnotesize\bf{ACKNOWLEDGMENTS}}
\end{center}
The authors acknowledge financial support from the National Research Fund Luxembourg under Grant No. CORE C22/MS/17415246/DeQuSky, the Icelandic Research Fund (Grant No. 217750), the University of Iceland Research Fund (Grant No. 15673), the Swedish Research Council (Grant No. 2020-05110), the Crafoord Foundation (Grant No. 20231063), and the European Research Council (ERC) under the European Union's Horizon 2020 research and innovation program (Grant No.\ 856538, project ``3D MAGiC'').









\clearpage
\newpage

\vspace{0.25cm}
\noindent
\textbf{Supplementary Note 1 $|$ Derivation of velocity equations}

Employing the spherical angles $(\Theta, \Phi)$, the magnetization vector of unit length can be parametrized $\mathbf{n}=(\sin\Theta\cos\Phi,\sin\Theta\sin\Phi,\cos\Theta)$.
In this case, the Landau-Lifshitz equation~\cite{Landau_Lifshitz} takes the following form,  
\begin{equation}
  \partial_{t}\Phi\sin\Theta=-\dfrac{\gamma}{M_\mathrm{s}}\dfrac{\delta\mathcal{E}}{\delta\Theta},\,\, \partial_{t}\Theta\sin\Theta=\dfrac{\gamma}{M_\mathrm{s}}\dfrac{\delta\mathcal{E}}{\delta\Phi},\label{LLG}
\end{equation}
where $\gamma$ is the gyromagnetic ratio.
At $t=0$ the system is in equilibrium, and the magnetization is given by $\Phi_{0}(\mathbf{r})=\Phi(\mathbf{r},0)$, $\Theta_{0}(\mathbf{r})=\Theta(\mathbf{r},0)$ satisfying the corresponding Euler-Lagrange equations. 
Assuming that the amplitude of the pulse is small, $\delta B(t)\ll B_{0}$, and neglecting the solitons' shape deformation, we obtain the following equations from \eqref{LLG}:
\begin{eqnarray}
    & \partial_{t}\Phi\sin\Theta_0= \gamma\delta B\sin\Theta_0, \,\,\partial_{t}\Theta\sin\Theta_0=0.
    \label{Phi_eq}
\end{eqnarray}
In the following, we consider two types of dynamics: translational motion and rotational motion of skyrmions.

\textbf{Translational motion.}
\noindent
For skyrmions moving with a constant velocity $\boldsymbol{v}$, meaning $\Phi(\mathbf{r},t)=\Phi(\mathbf{r}-\boldsymbol{v}t)$ and  $\Theta(\mathbf{r},t)=\Theta(\mathbf{r}-\boldsymbol{v}t)$, we get from \eqref{Phi_eq} the following equation:
\begin{equation}
    -\boldsymbol{v}\cdot\nabla\Phi_{0} =\gamma\delta B.
    \label{v_eq1_F}
\end{equation}
Multiplying \eqref{v_eq1_F} by magnon density $\mu=(4\pi)^{-1}(1-\cos\Theta_0)$ and performing the integration we obtain 
\begin{equation}
    \boldsymbol{v}\cdot\boldsymbol{p}=-\gamma m \delta B,
    \label{v_eq2}
\end{equation}
where $\boldsymbol{p}$ and $m$ are momentum and magnonic number defined as follows:
\begin{equation}    \boldsymbol{p}=\displaystyle\int\mu\nabla\Phi_0\mathrm{d}S,\,\,m = \int \mu\mathrm{d}S.
\label{magnonic_number}
\end{equation}
As shown in Ref.~\cite{Papanicolaou}, for a translation-invariant Hamiltonian, $\boldsymbol{p}$ and $m$ remain conserved quantities, even when the profiles of $\Theta$ and $\Phi$ deviate from the static configuration.

The magnonic number $m$ is invariant under rotations and translations of the texture, but this is not true for the momentum. 
Under rotation of the skyrmion about $z$-axis, $\boldsymbol{p}$ rotates.
Note, momentum $\boldsymbol{p}$ is translation-invariant only if $Q=0$.
Thus, $\mathbf{v}=0$ for any soliton with $Q\neq 0$.
Equation \eqref{v_eq2} has non-trivial solutions only if $Q=0$ and $\boldsymbol{p}\neq 0$.
The latter implies that $k_\mathrm{s}=1$. 
Therefore, all skyrmions that exhibit translational motion under an external field pulse should be topologically trivial and should have low symmetry.

We can write the velocity vector in terms of the speed $v=|\boldsymbol{v}|$ and deflection angle $\psi$: $(v_\mathrm{x},v_\mathrm{y})=v(\cos\psi,\sin\psi)$. 
To determine $v$ and $\psi$, we have to consider \eqref{v_eq2} and the second equation in \eqref{Phi_eq}.
Substituting in the latter the rigid translation ansatz gives:
\begin{equation}
-\boldsymbol{v}\cdot\nabla\Theta_{0}\sin\Theta_0=0.
\label{v_eq4}
\end{equation}
Then we can multiply \eqref{v_eq4} by $\lvert \mathbf{r} - \mathbf{r}_{0}\rvert^2$, where $\mathbf{r}_{0}$ is the skyrmion guiding center 
and integrate by parts over $xy$-plane:
\begin{equation}
\boldsymbol{v}\cdot\int (\mathbf{r}-\mathbf{r}_{0})\mu \mathrm{d}S=v_\mathrm{x}m_\mathrm{y}-v_\mathrm{y}m_\mathrm{x}=0,
\label{v_eq5}
\end{equation}
where $m_\mathrm{x}$, $m_\mathrm{y}$ are first moments of the magnon density \eqref{magnonic_number}.
The corresponding vector $\boldsymbol{m}=(m_\mathrm{x}$, $m_\mathrm{y})$ transforms in the same way as $\boldsymbol{p}$ under rotation, and are invariant under translation.
Thus, we can use \eqref{v_eq5} to find the deflection angle, $\tan\psi=m_\mathrm{y}/m_\mathrm{x}$, up to a $\psi \mapsto \psi+\pi$ ambiguity. 
This means we know the line of motion but not the direction of the soliton along it.

Then, the speed of the skyrmion follows from \eqref{v_eq2}:
\begin{eqnarray}
v =  \dfrac{\BBS{-}\gamma m \delta B}{p_\mathrm{x}\cos\psi+p_\mathrm{y}\sin\psi}.
    \label{v_sol}
\end{eqnarray}
The ambiguity in $\psi$ can equally be seen as an ambiguity of sign in $v$. 
Note that as we change the sign of $\delta B$, we flip the direction of motion, but we cannot predict which sign of $\delta B$ will correspond to which direction.
\begin{figure}[htb!]
\centering
\includegraphics[width=8.1cm]{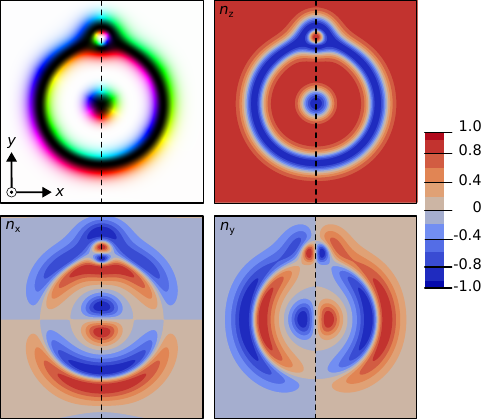}
\caption{\small ~\textbf{Skyrmion mirror symmetry}.
Skyrmion bag with $Q=0$, $k_\mathrm{s}=1$ and its magnetization components ($n_\mathrm{x}$, $n_\mathrm{y}$, $n_\mathrm{z}$) are shown.
The vertical dashed line corresponds to the skyrmion symmetry axis.
}
\label{Fig3}
\end{figure}

Further simplification of \eqref{v_sol} can be made for solitons having an additional mirror symmetry. 
To exemplify this symmetry, we consider the Snowman skyrmion shown in Fig.~\ref{Fig3}.
As one can notice, magnetization components are characterized by additional reflection symmetry: 
$n_\mathrm{x}(x,y)=n_\mathrm{x}(-x,y)$, $n_\mathrm{y}(x,y)=-n_\mathrm{y}(-x,y)$, and $n_\mathrm{z}(x,y)=n_\mathrm{z}(-x,y)$.
The symmetry axis, which coincides with the $y$-axis, so $\psi = 0$.
Thus, in the example pictured the direction of motion is along the $x$-axis.
For symmetry axis oriented at an arbitrary angle $\psi_0$ to the $x$-axis, $\boldsymbol{m}$ and $\boldsymbol{p}$ are both perpendicular to the line of symmetry, meaning $\psi=\psi_0\pm\frac{\pi}{2}$.
It turns out that this mirror symmetry is widespread among skyrmions. 
When it is not true, the general formula \eqref{v_sol} has to be used.

Interestingly, in the case of translational dynamics of mirror-symmetric skyrmions, we do not have $\boldsymbol{p}\propto\boldsymbol{v}$ like in Newtonian mechanics, but instead an inverse relation.
This indicates a significant difference between momentum in Newtonian mechanics and in the theory of magnetic solitons.
Nevertheless, $\boldsymbol{v}$ is parallel to $\boldsymbol{p}$, in accordance with numerical observations, as we demonstrate below.

\textbf{Rotating motion.}
\noindent
In the case of skyrmion rotation with angular velocity $\omega$, the magnetic texture motion is described as $\Phi(x,y,t)=\Phi(x^\prime,y^\prime)-\omega t$ and  $\Theta(x,y,t)=\Theta(x^\prime,y^\prime)$ with rotated coordinates:
\begin{eqnarray}
	& x^{\prime}=x\cos \omega t-y\sin\omega t,\,\, y^{\prime}=y\cos \omega t+x\sin \omega t.
	\label{rot_coord}
\end{eqnarray}
where $\omega$ is the angular velocity of skyrmion. So the derivative with respect to time can be found as
\begin{equation}
    \dfrac{\partial\Phi}{\partial t}=-\omega-\omega\left(y\dfrac{\partial\Phi}{\partial x}-x\dfrac{\partial\Phi}{\partial y}\right).\label{dFt_w}
\end{equation}
Then substituting this into \eqref{Phi_eq}, we get the following equation
\begin{equation}
    -\omega-\omega\left(y\dfrac{\partial\Phi_0}{\partial x}-x\dfrac{\partial\Phi_0}{\partial y}\right)=\gamma\delta B.\label{w_eq1}
\end{equation}
Multiplying \eqref{w_eq1} by $\mu$ and performing integration lead to
\begin{equation}
   \omega m + \omega\displaystyle\int\left(y\dfrac{\partial\Phi_{0}}{\partial x}-x\dfrac{\partial\Phi_0}{\partial y}\right)\mu\mathrm{d}S=-\gamma m \delta B.\label{w_eq2}
\end{equation}
The integral in \eqref{w_eq2} is nothing but angular momentum, $l$, employing which, the solution for the angular velocity can be found from \eqref{w_eq2}:
\begin{equation}
    \omega = \dfrac{\gamma m\delta B}{l-m}.\label{w_sol}
\end{equation}
From \eqref{second_top_moment} follows that $l$ takes non-zero values for skyrmions of arbitrary $k_\mathrm{s}$. However, the Thiele equation has nontrivial solutions, $\omega\neq0$, only for skyrmions with $k_\mathrm{s}>1$.

To use the formula \eqref{v_sol}, \eqref{w_sol}, one must compute the corresponding momentum integrals.
However, this might represent a challenge due to the non-differentiability of $\Phi$ at skyrmion cores, $\Theta=0$ and $\Theta=\pi$.

\textbf{Supplementary Note 2 $|$ Calculation of momentum integrals}.
\noindent
The momentum $\boldsymbol{p}=(p_\mathrm{x},p_\mathrm{y})$ \eqref{magnonic_number} can be rewritten in terms of the topological density:
\begin{eqnarray}
    & p_\mathrm{x}= -\displaystyle\int q y \mathrm{d}S + \displaystyle\int \mu y\varepsilon_{ij}\partial_{i}\partial_{j}\Phi_0\mathrm{d}S, \nonumber \\
    & p_\mathrm{y}=\displaystyle\int q x \mathrm{d}S-\displaystyle\int \mu x\varepsilon_{ij}\partial_{i}\partial_{j}\Phi_0\mathrm{d}S, \label{top_moment}
\end{eqnarray}
where $\varepsilon_\mathrm{xy}=-\varepsilon_\mathrm{yx}=1$ and $\varepsilon_\mathrm{xx}=\varepsilon_\mathrm{yy}=0$. 

The method for computing the last integrals in \eqref{top_moment} depends on the type of cores the particular skyrmion has.
An important characteristic here is the winding number:
\begin{equation}
    \nu = \lim_{\epsilon\rightarrow0} \dfrac{1}{2\pi}\oint_{C_\epsilon} \nabla\Phi\cdot\mathrm{d}\mathbf{l},\label{winding}
\end{equation}
where the integration contour $C_\epsilon$ is defined by $\Theta(\mathbf{r})=\pi-\epsilon$. 

Some skyrmions can have a few cores, as shown in Fig.(2) \textbf{b}, \textbf{c} in the main text.
While axially symmetric $\pi$-skyrmion \textbf{a} has only one core, chiral droplet \textbf{b} and antiskyrmion \textbf{c} have two and three cores, respectively. 
These multicore situations are quite generic, and more complex solitons, as a rule, are characterized by more cores.
In some cases, a set of points with $\Theta(\mathbf{r})=\pi$ can represent a curve, as for skyrmionium and other skyrmion bags.
In the former and the latter cases, we will refer to such cores as simple cores and complex cores, respectively.

In the case when skyrmions have simple cores only,
the last integrals in \eqref{top_moment} are proportional to the delta function taken at the core of solitons $\Theta=\pi$ and multiplied by the corresponding vector field vorticity, $\nu_{i}$, around the core:
\begin{equation}
    p_\mathrm{x}= -\displaystyle\int q y \mathrm{d}S + \sum_{i} \nu_{i} y_{i}, \,\,p_\mathrm{y}=
    \displaystyle\int q x \mathrm{d}S- \sum_{i} \nu_{i} x_{i},
    \label{top_moment_2}
\end{equation}
where $\mathbf{r}_{i}=(x_{i},y_{i})$ is the $i$-th core of the skyrmion.
It is from this expression that we can see that under a translation $\mathbf{r}\mapsto \mathbf{r}+(a_\mathrm{x},a_\mathrm{y})$, the contributions from the integral and the sum are equal and one has $\boldsymbol{p}\mapsto\boldsymbol{p} +8\pi Q (-a_{y}, a_{x})$, and thus $\boldsymbol{p}$ is only translation-invariant if $Q=0$. 
This statement was made more generally in Ref.~\cite{Papanicolaou}, but without taking into account the delta-function contributions.
Similarly, the angular momentum, $l$, can be calculated employing the topological charge density as
\begin{equation}
 l = -\dfrac{1}{2}\displaystyle\int q r^{2}\mathrm{d}S+\dfrac{1}{2}\sum_{i} \nu_{i}r_{i}^{2}.
 \label{second_top_moment}
\end{equation}

Complex cores can be split into two cases. In the first case, when the set of points with $\Theta(\mathbf{r})=\pi$ forms a single closed curve, e.g. skyrmionium, we can always define $\Theta>\pi$ on one side $\Theta<\pi$ on the other so that $\Phi$ has no singularities, and there is thus no contribution to the integrals above. We can say these cores have no winding. After removing such cores, we have the second case, where $\Theta=\pi$ forms a curve with ends and $\Phi$ may be discontinuous across the curve. In this case the winding can be defined as in \eqref{winding} and it will in general be non-zero.

This situation is exemplified by Fig. \ref{Fig3}, where there are three curves where $\Theta=\pi$ forming a `figure-eight' shape. By noting that $\Phi$ increases anticlockwise along each of these curves, we see that the outer curve can be seen as a zero-winding core, and the central bridge separating the two closed regions is an extended core with non-zero winding.

While the formula for the contribution of these cores to the winding is more complicated, it has the same behavior under translation as simple cores. This means that again $\boldsymbol{v}=0$ for any $Q\neq0$ texture.

\appendix
\onecolumngrid

\end{document}